\documentclass[]{aastex631}
\usepackage{csquotes,hyperref,pgffor}

\received{21 March 2022}
\revised{27 May 2022}
\accepted{6 June 2022}
\submitjournal{Acta Astronautica}

\shorttitle{SETI in 2021}
\shortauthors{Huston \& Wright}
\graphicspath{{./}{figures/}}
\newcommand{\PSUAA}{Department of Astronomy \& Astrophysics, Pennsylvania State University, University Park, PA, 16802, USA}
\newcommand{\PSUCEHW}{Center for Exoplanets and Habitable Worlds,  Pennsylvania State University, University Park, PA, 16802, USA}
\newcommand{\PSETI}{Penn State Extraterrestrial Intelligence Center, Pennsylvania State University, University Park, PA, 16802, USA}

\newcommand{\mycounter}[2]{
\newcounter{#1}
\setcounter{#1}{#2}
}

\renewcommand{\th}{\textsuperscript{th}}

\mycounter{Npapers}{0}
\mycounter{searches}{14}
\addtocounter{Npapers}{\thesearches}
\mycounter{methods}{14}
\addtocounter{Npapers}{\themethods}
\mycounter{targets}{13}
\addtocounter{Npapers}{\thetargets}
\mycounter{development}{12}
\addtocounter{Npapers}{\thedevelopment}
\mycounter{theory}{17}
\addtocounter{Npapers}{\thetheory}
\mycounter{social}{28}
\addtocounter{Npapers}{\thesocial}

\begin{document}

\title{SETI in 2021}

\correspondingauthor{Macy Huston}
\email{mhuston@psu.edu}

\author[0000-0003-4591-3201]{Macy J.\ Huston}
\affil{\PSUAA}
\affil{\PSUCEHW}
\affil{\PSETI}

\author[0000-0001-6160-5888]{Jason T.\ Wright}
\affil{\PSUAA}
\affil{\PSUCEHW}
\affil{\PSETI}

\begin{abstract}

In this second installment of SETI in 20xx, we very briefly and subjectively review developments in SETI in 2021. Our primary focus is \theNpapers\ papers and books published or made public in 2021, which we sort into six broad categories: results from actual searches, new search methods and instrumentation, target and frequency selection, the development of technosignatures, theory of ETIs, and social aspects of SETI. 

\end{abstract}


\section{Introduction} \label{sec:intro}

This is the second installment of the SETI in 20xx a series, begun by \citet{SETIin2020}, providing a review of papers in SETI from the previous year. This series is inspired by the "ApXX" annual reviews of all of astrophysics by Virginia Trimble \citep[e.g.][]{Trimble2005}, but is considerably narrower in scope.

SETI is a broad field, including technosignature searches, the theory that underlies it, and associated topics such as METI. To help constrain the scope of this series, we start with the ADS SETI bibliography \citep{Reyes2019,LaFond2021} to find all of the works with publication dates in 2021 in the field as defined in those works. We do not claim or even attempt to make our listing here complete, and our decisions about the scope of this paper may strike some as idiosyncratic. In the spirit of Trimble's reviews, our aim is to be ``usefully subjective.'' 

For the most part this means we have selected refereed papers published in their final form in 2021 that are firmly, and not merely incidentally or tangentially, about SETI, but we also included some other entries where it seemed appropriate to us. As in 2020, we have avoided compilations of previously published work, and saved some preprints appearing at the end of the year for the next installment, once they have a final bibliographic citation. 

To go beyond our review, we invite readers to use the ADS \texttt{bibgroup:SETI} and \texttt{year:2021} search query elements\footnote{That is, enter  \href{https://ui.adsabs.harvard.edu/search/q=bibgroup\%3ASETI\%20year\%3A2021&sort=date\%20desc\%2C\%20bibcode\%20desc&p_=0}{bibgroup:SETI year:2021} in the ADS URL search tool. As a guide, we provide hyperlinks to this search and others throughout this paper}. This will produce a more comprehensive survey of scientific output in the field in 2021, and one that is more objective, because that bibgroup is maintained according to the methodology described in \citep{Reyes2019,LaFond2021}.

We have striven to roughly track the scope and structure of the previous entry in this series, and anticipate that we will do the same in future editions. This means we have once again roughly categorized these papers into six categories: results from actual searches, new search methods and instrumentation, target and frequency selection, the development of technosignatures, theory of ETIs, and social aspects of SETI (including matters concerning METI).

\section{SETI in 2021}

The forecast in our previous installment for prospects for in-person meetings in 2021 proved to be too optimistic, and many opportunities for SETI practitioners to meet and interact were lost again to the COVID pandemic in 2021, especially with the impact of the Delta and Omicron variants. For instance, the First Penn State SETI Symposium was postponed \textit{again}, this time to June 2022. 

However, the SETI community continued to find ways to interact through virtual meetings. The Penn State Extraterrestrial Intelligence Center sponsored the 2021 Assembly of the Order of the Octopus\footnote{\href{https://zenodo.org/communities/octopusassembly2021/?page=1\&size=20}{Zenodo: octopusassembly2021}}, which was held virtually, and served as the first in a planned series of conferences for early-career researchers in SETI. Six SETI abstracts were featured in the virtual 237\th\ Meeting of the American Astronomical Society\footnote{\href{https://ui.adsabs.harvard.edu/search/q=docs(library\%2Fk1BwfM56QgKbl6X-PXADqg)\%2C\%20\%20year\%3A2021\%2C\%20\%20bibstem\%3A\%22aas\%22&sort=date\%20desc\%2C\%20bibcode\%20desc\&p_=0}{ADS query: bibgroup:SETI,  year:2021,  bibstem:``aas''}}, along with 6 more at the virtual Committee on Space Research Scientific Assembly\footnote{\href{https://ui.adsabs.harvard.edu/search/q=docs(library\%2Fk1BwfM56QgKbl6X-PXADqg)\%2C\%20\%20year\%3A2021\%2C\%20\%20bibstem\%3A\%22cospar\%22&sort=date\%20desc\%2C\%20bibcode\%20desc&p_=0}{ADS query: bibgroup:SETI,  year:2021,  bibstem:``cosp''}}.

2021 was a year of growth for SETI education. Penn State's first undergraduate SETI course was held, complementing its existing graduate course, as was the sixth instance of the graduate/undergraduate SETI course at UCLA. The year also saw two PhD graduations with SETI-focused theses: Dr. Sofia Sheikh of Penn State \citep[][advised by Jason Wright]{Sheikh2021d} and Dr. Paul Pinchuk of UCLA \citep[][advised by Jean-Luc Margot]{Pinchuk2021b}. \citet{LingamTextbook} have published a textbook \textit{Life in the Cosmos: From Biosignatures to Technosignatures} which includes an extensive, upper-level undergraduate or graduate-level discussion of SETI. 

This year saw a significant increase in SETI papers from 2020's 75 papers \cite{SETIin2020} to \theNpapers. While Results from Searches in 2020 comprised 7 papers, primarily focused on radio frequencies, 2021 saw 14 papers, including searches for infrared waste heat and artifacts among more radio results. We see similar output for Search Methods and Instrumentation in both years, with continuing work in radio technology specifically for SETI, the application of other extant telescopes to SETI, and the development of anomaly detection methods. 2021 saw in increase in papers about Target and Frequency Selection, continuing the exploration of possible Schelling points and assessing the technological habitability of possible targets. The Development of Technosignatures remained a topic of interest in 2021, with 12 papers further exploring technical aspects of spacecraft and and megastructures, as well as the presentation of a few new potential technosignatures. We again see a similar level of output in the Theory of ETIs, continuing conversations around the Drake Equation, Fermi Paradox, Copernican and anthropic principles, and the evolution of exocivilizations. 2021 saw a staggering 28 new papers about Social aspects of SETI, compared to last year's 10. This is in part due to the special American Indian Culture and Research Journal issue, with 9 articles  focusing on SETI and related topics. As in 2020, this category also contained historical work and discussions of METI and contact.
 
Also in 2021, Dan Werthimer and Paul Horowitz shared the SETI Institute's Drake Award (which was not awarded in 2020).  These are well deserved honors: Paul ran the META and BETA radio SETI searches, has advised nearly half of all PhD thesis in the field, pioneered pulsed laser SETI, and has been instrumental in PanoSETI.  Dan has run the CASPER and SERENDIP projects to develop radio SETI hardware, and served as chief scientist of the groundbreaking and extremely popular SETI@HOME citizen science project.
 
Finally, the SETI community sadly lost at least three prominent members in 2021. Bob Gray, who called himself an ``amateur'' astronomer, did foundational work on the Wow! Signal, and contributed to our understanding of the history of the field (his final work in the field appears below). Sasha Zaitsev coined the term ``METI'' and wrote about its rationale and the ``SETI Paradox.'' Lev Gindilis was one of the founders of the field. He attended the 1964 Byurakan conference on Extraterrestrial Civilizations and published across 5 decades, most recently a history of SETI in Russia \citep{Gindilis2019}.

\section{Results from Searches (\thesearches\ papers)}

The heart of SETI is the actual searches, and 2021 brought several new important results and upper limits.

The most publicised SETI event this year was certainly the Breakthrough Listen (BL) signal of interest blc1, producing three papers from the collaboration and many articles in the media. \citet{Smith2021b} described the BL search for radio technosignatures from the direction of Proxima Centauri using the Parkes Murriyang radio telescope and the detection of blc1. Its companion paper, \citet{Sheikh2021b},  presented the analysis of the technosignature candidate, which determined that the signal was not extraterrestrial, and instead produced by human-made electronics. Follow-up observations of Proxima Centauri by \citet{Sheikh2021a} found no repeats of blc1 or any other signals of interest near the same wavelength. 

The BL team also published two other radio technosignature searches with no candidates found, setting upper limits on the presence of radio transmitters in the observed space. \citet{Gajjar2021} presented their observing strategy and results of their search for radio technosignatures pointed near the Galactic center, a line of sight which offers the greatest number of potentially habitable systems to be observed at once. \citet{Traas2021} targeted 28 stars identified by the Transiting Exoplanet Survey Satellite (TESS) as possible exoplanet hosts.

Also in the radio, \citet{Margot2021} searched for radio technosignatures in the regions surrounding 31 Sun-like stars near the Galactic plane, using an improved candidate signal detection procedure, finding only RFI. \citet{Crilly2021a, Crilly2021b} described the hypothesis behind and performed a novel technosignature search for radio communication via polarized pulse pairs. \citet{Montegubnoli21} presented the history of the Medicina radio telescope's SETI searches and its prospects for the future, which were also discussed by \citet{lulli21}.

In the infrared, \citet{Chen2021a} searched for Kardashev Type III civilizations which generate waste heat from Dyson sphere-like structures. They used the LoTSS-DR1 value-added catalog to study the infared-radio correlation of $\sim$16,000 galaxies, finding two galaxies to be Type III civilization candidates for further study and noting the value of extending this analysis to the full LoTSS survey to place upper limits on Type III civilizations in the universe.

%Closer to home, \citet{Gillon2021} explored the hypothesis that self-reproducing probes formed an interstellar communication network via stellar gravitational lensing. Targeting the Sun's focal line with respect to Wolf 359, our third nearest star system, as well as one which can see Earth transit, they detected no outgoing optical signals and no object consistent with the hypothesized transmitter motion within 20 AU.

Conversations around the interstellar object 'Oumuamua which passed through the solar system in 2017 continued in 2021. \citet{Zuckerman2021} demonstrated that the capabilities of large space telescopes are superior to any spaceship resembling 'Oumuamua, indicating that there would be no motivation for sending such a spaceship. \citet{Curran2021} also rejected the hypothesis that 'Oumuamua was of artificial origin, citing technical concerns about light sails. 
%\citet{Loeb2021} addressed some proposed natural explanations for the object and insisted that the possibility of artificial origin remain in consideration. 
\citet{Cowie2021} discussed the philosophical flaws in the argument that 'Oumuamua is an extraterrestrial artifact. %\citet{Elvis2021} discussed interstellar travel and research programs inspired by the artificial origin hypothesis, e.g. missions to intercept interstellar objects discovered in the future.

\section{Search Methods and Instrumentation (\themethods\ papers)}

\subsection{Instrumentation}

In work originating from the 2020 TechnoClimes workshop, \citet{Socas-Navarro2021} presented a variety of possible technosignature search mission concepts. They also introduced the very useful ``ichnoscale'' to quantify technosignatures with respect to current Earth technology (normalized so that the technosignatures of the Earth today have an ichnoscale score of $\iota = 1$.)

Other work described new SETI-specific hardware on extant telescopes. \citet{Price2021} described the new ultra-wideband low (UWL) receiver for the Parkes telescope and updates to their data recording hardware for new technosignature observations. \citet{Murphy2021} reported on the techniques and first results for the REAL-time Transient Acquisition backend at the Irish LOFAR station, including the hardware installed for Breakthrough Listen SETI searches. \citet{melis21} presented plans for the Sardinia Radio Telescope to conduct SETI observations, as part of the Breakthrough Listen initiatives. 

\citet{Chen2021b} assessed the application of the finite Koch snowflake antenna on the Five-hundred-meter Aperture Spherical radio Telescope (FAST) radio telescope to optimize SETI searches and potentially detect Kardashev Type-I civilizations. \citet{schillir21} presented commensal SETI observing and data analysis strategies for the upcoming Square Kilometer Array. \citet{antoniettei21} described a new technique for the Karhunen Loève Transform, a data processing method for radio SETI searches.

The SETI capabilities of extant instrumentation were also discussed. \citet{Panov2021} described the TAIGA-HiSCORE Cherenkov array's search for nanosecond optical transients, including its application to SETI with a very wide-field of view optimal for detecting very rare events. \citet{Caraveo21} discussed the use of Cherenkov telescopes for optical SETI, as commensal observations taken in parallel with gamma ray searches.

\citet{Osmanov2021a} concluded that Galactic and extragalactic self-replicating von-Neuman probes might be detectable with FAST. \citet{Osmanov2021b} assessed the capability of 1-meter-scale optical telescope to detect hot megastructures around stars, both via thermal emission and optical variability. 

\subsection{Anomaly detection}

As part of the Vanishing \& Appearing Sources during a Century of Observations (VASCO) project, \citet{Villarroel2021b} reported on nine point sources which appeared on  a photographic plate in the Palomar Sky Survey in April 1950 and neither of the surrounding images, nor deeper modern surveys. They explored several possible human-caused and natural explanations, concluding that radioactive bomb particles may be the most down-to-Earth explanation but further study is worthwhile. %\citet{Villarroel2021b} present a strategy to find Non-Terrestrial artifacts in geosynchronous orbit as glints in pre-satellite imaging data.

\citet{Kipping2021a} put forth a general statistical framework for interpreting rare ``black swan'' astronomical events and guiding future observations. \citet{Martinez-Galarza2021} developed a method to identify, rank, and categorize anomalous events in time-series data sets.

\section{Target and Frequency Selection (\thetargets\ papers)}

\citet{Wright2021a} provided an overview of SETI strategies and advice for those involved in the field. \citet{George2021} interpreted SETI as one-shot hypothesis testing, where any detection is either a natural process or sent from ETI, and established a mathematical framework to optimize methods for sending and receiving interstellar messages.

\citet{Czech2021} presented Breakthrough Listen's target selection strategy for commensal SETI observations on the MeerKAT telescope, which aims to observe a million stars of well-understood types from the Gaia DR2 database over $\sim$13 months. \citet{Houston2021} described a methodology to optimize the expected ET detection rate through the joint consideration of field of view and sensitivity. \citet{Denissenya2021} discussed applications of Ultra Fast Astronomy optimization strategies to optical SETI, focusing on the detection of millisecond to nanosecond timescale flashes.

Following \citet{Seto19}, \citet{Seto2021} further explored the idea of communication timing based around the Schelling points of supernovae in the Galaxy, identifying an optimal time after reference burst observation to send or search for artificial signals. 
On the question of where to look, \citet{Kaltenegger2021} suggested studying stars who could detect Earth and identify it as a possible site for life. They identified nearby ($\lesssim$100 pc) stars which could have seen Earth transit since early human civilization has existed, of which 75 are near enough for humans' artificial radio leakage to have reached. \citet{vladilo} discussed the use of climate models to predict which rocky exoplanets may be habitable for multicellular life, as a method to identify SETI search targets.

Other work focused on the prospects for technological life in different regions and types of Galaxies. \citet{Smith2021c} examined the conditions (e.g. metallicity and stellar density) which may make globular clusters favorable for an extraterrestrial civilization, identifying NGC 6553 as an ideal candidate. \citet{Wright2021b} and \citet{Lares2021} focused on the impact of galactocentric distance on extraterrestrial civilizations, both concluding that the central (densest) regions of galaxies are optimal. However, \citet{Lacki2021c} found that the same is not true in elliptical galaxies as it is for spirals, because the inner $\sim$hundred parsec zones would typically be too collisionally hazardous for planets to develop intelligent life. On which types of galaxies are optimal for SETI searches, \citet{Lacki2021b} introduces the concept of ``galactic traversability'' as a limiting factor in the development of galactic-scale societies, concluding that compact galaxies with minimal interstellar medium may be optimal extragalactic SETI targets.

\section{Development of Technosignatures (\thedevelopment\ papers)}

\subsection{Dyson spheres  and spacecraft}
Several recent works have explored the interstellar communication network hypothesis, which proposes the use of stars as gravitational lenses to amplify interstellar signals from probes in neighboring star systems. \citet{Maccone2021a} explored the information channel capacity of such radio bridges, examining the bit error rate for communication to nearby stars via the Sun's gravitational lens, as well as a radio bridge which uses not only the Sun as a gravitational lens but also the receiver system's star. \citet{Hippke2021b} set forth the third work in the ``Interstellar communication network'' series, describing how to locate candidate sky positions for nodes in the solar system. \citet{Kerby2021} examined the engineering requirements of such a stellar relay system, considering how stellar and planetary system architectures affect the ability to maintain relay-star-target alignment. \citet{Gertz2021a} elaborated on how a Galactic communication network may operate and concluded that the search for such probes may be more fruitful than the search for electromagnetic signals targeted toward Earth.

\citet{Osmanov2021c} extrapolated Earth's satellite megaconstellations to consider the detectability of a Kardashev Type-I society, with a megastructure surrounding a habitable zone Earth-like planet. They determined that the Very Large Telescope Interferometer could detect such a structure in the star's IR spectrum, hypothesized that spectral variation may be able to detect the megastructure's motions, and argued that FAST may be able to detect such structures if they emit in the radio. \citet{Osmanov2021d} proposed that the Fermi paradox may be resolved by the existence of many extraterrestrial civilizations hosting ring-like megastructures which produce signals similar to those from natural stellar oscillations and thus go unnoticed by observers.

\citet{Almeida2021} discussed Roger Penrose's proposal of a mechanism to extract rotational energy from a Kerr black hole in 1969. \citet{Hsiao2021} explored various mechanisms by which a Dyson sphere around a black hole could power a Kardashev Type-II civilization, collecting radiation due to accretion as well as kinetic energy from jets, and described how such a structure might appear in spectral energy distributions.

\subsection{Atmospheric and Geological Technosignatures}

\citet{Lockley2021a} considered the detectability of pre-industrial societies, with a focus on the signs of seasonal agriculture, also mentioning potential detectability for urban fires, land-use or aquatic change, introduced species, and climate change. \citet{Savitch2021} explored the coupled evolution of a civilization harvesting energy and its planet, which undergoes climate change. They found that an ``anthropocene'' like that on Earth can be triggered by combustion creating CO$_2$ in most planets in their parameter space for initial chemistry and residing within different parts of the Complex Life Habitable Zone. 
\citet{Kopparapu2021} modelled the spectral signature of NO$_2$, an industrial byproduct, in the atmospheres of exoplanets, finding that a LUVOIR-like telescope would require several hundreds of hours of observation time to detect present Earth-level NO$_2$ on a nearby exoplanet. 
%\citet{Beatty2021} explore the detectability of city lights on the night side of exoplanets, from Earth-like levels to fully urbanized planets. They conclude that several nearby exoplanets' city lights would be detectable with a few hundred hours of LUVIOR observation at 3$\sigma$ for urbanization levels near Earth's, and planet-wide cities, or ``ecumenopolis,'' would be detectable around several tens of nearby stars.

\subsection{Other novel technosignatures}

\citet{Hippke2021a} proposed a search for interstellar quantum communications and described how current technology may be used to detect Fock state photons or squeezed light, which would be unambiguously artificial.

\section{Theory of ETIs (\thetheory\ papers)}

%Theory on the nature, scale, lifetime, and ubiquity of ETIs was one of the largest categories in 2021, something I suspect is true most years.

\subsection{The Drake Equation \& Fermi Paradox}

The Drake Equation, its meaning, shortcomings, and estimates of its included variables, often dominates the conversation in SETI theory, and it played a large role in the conversation again in 2021. 
\citet{Golden2021} estimated the product of the Drake equation's non-longevity factors as random variables, using Monte Carlo analysis to find a value of 0.85 $\pm$ 1.28 years$^{-1}$, which is in close agreement with Drake's estimate and that of the Project Cyclops report.

Some work sought to reformulate the Drake equation, bringing in additional considerations. \citet{Kipping2021b} presented a stochastic formulation of the Drake equation as a Poisson process dependent only on the mean rate of civilization births in the Milky Way ($\lambda_C$) and a rate of collapse ($\lambda_L$), also applying the formulation to the Copernican Principle and the probability we are alone in the Galaxy. \citet{Smith2021d} examined the effect of relationships between ETIs on civilization emergence and lifetime, considering how the Drake equation expectations change depending on how beneficial or detrimental the interactions are.

Another common theme in Drake equation literature is reformulating the equation to apply it to different aspects of ETI and technology. \citet{Grimaldi2021} explored two new parameters related to the Drake equation to calculate the number of communicative electromagnetic signals present ($N_G$) in the Milky Way and the average number of them which cross Earth ($\overline{k}$), which are functions of emission birth rates, longevities, and directionality. \citet{Benford2021} proposed a version of the Drake equation which includes the search for extraterrestrial artifacts in the solar system (SETA), comparing the strategies and chances of success for SETI and SETA. 

Lastly, some work sought to fully replace or abandon the equation. \citet{Cai2021} examined the occurrence rate of ETI in the Milky Way with an alternative equation which considers three main parameters: the likelihood rate of abiogenesis, evolutionary timescales, and the probability of self-annihilation of complex life. \citet{Gertz2021b} ``abandoned'' the Drake equation based on its inability to provide a precise answer to the prevalence of ETI, concluding that such a question can only be answered with observations.

Related to the Drake Equation is the Fermi Paradox, or lack of ETI detections despite estimates that they are common. \citet{SLIJEPCEVIC2021} argued that this is due to the focus on humanoid intelligences and communication methods. They described intelligence or cognition existing in ensembles of bacteria, proposing that this form of intelligence may be much more common in the universe and spread via panspermia.

\subsection{The Copernican and Anthropic Principles}

The Copernican and Anthropic principles involve similar trains of thought as the Drake equation and Fermi Paradox, with a particular focus on the implications of humanity's existence. \citet{Siraj2021b} applied the Copernican principle, which asserts that humans are not privileged observers in the universe, to blc1, concluding that a radio-transmitting civilization around a Sun-like star so closed to us is exceptionally improbable. \citet{Siraj2021a} also applied the Copernican framework to the question of when we may receive a response from ETI who have detected radio transmissions from Earth, concluding that such a response should not be expected for millennia.

\citet{Lacki2021d} presented the Fine Graining with Auxilary Indexicals framework to analyze typicality arguments and its application to the Doomsday Argument, a controversial application of the Copernican principle that concludes that it is extremely unlikely humanity has a future with many more humans than have existed in Earth's past. \citet{balbi21} described the issues involved in assuming that we are in a typical cosmological epoch.

\subsection{The Evolution and Expansion of ETIs}

Literature discussing the evolution of complex or intelligent life often invokes the ``hard steps'' model, which characterizes the development of such life as a series of major evolutionary transitions. \citet{Snyder-Beattie2021} explored the timings of key evolutionary transitions on Earth and concluded that the expected transition times likely exceed the lifetime of Earth, suggesting that extraterrestrial intelligent life may be extremely rare. \citet{Hanson2021} invoked hard steps in a simple model of ``grabby'' (or expansive) ETIs, which expand through their galaxy at a constant speed, arising from more common ``quiet'' civilizations, concluding that non-detection indicates that these quiet civilizations are rare to begin with or that the transition to grabby is exceedingly rare.

Perhaps the least well constrained factor in the Drake equation is the longevity of technological civilizations, $L$. \citet{Balbi2021} explained why the focus should be on the longevity of technosignatures, rather than of the life that produces them, as many types of technology may outlive biological life. They also argued that technosignature searches should focus on those which are long-lived as they are much more likely to be detected. \citet{Hansen2021} explored stellar flybys as an opportunity for shorter travel times in interstellar migration, necessitated by stellar evolution, and argued that this offers an explanation for the Fermi Paradox.

\citet{Ord2021} examined the effects of the universe's accelerating expansion on the limits of contact and travel for extra- and inter-galactic civilizations.

\section{Social aspects of SETI (\thesocial\ papers)}

\subsection{Indigenous Studies}
In December 2021, the American Indian Culture and Research Journal (AICRJ) published a special issue entitled ``Settler Science, Alien Contact, and Searches for Intelligence,'' with guest editors David Delgado Shorter and Kim TallBear. \citet{Shorter2021a} described how this volume evolved out of the Native Studies Working Group at a Breakthrough Listen workshop in 2018. BL posed the question: ``what would
you most want SETI scientists to know about potentially making contact?'' \citet{Atalay2021} presented the Indigenous Studies Working Group Statement in response to BL's question, recommending the creation of a Mission Statement to clearly answer ethical questions about why to search, whether the search can be harmful, whether ET life can refuse contact, and to acknowledge the non-objective nature of terms used in the field (e.g. ``intelligent,'' ``advanced,'' and even ``life''). 

Further articles in the AICRJ issue examined more broadly the ethics of contact and SETI, and other aspects of astronomy and space exploration. \citet{Shorter2021b} discussed the language and analogies used in SETI, emphasizing the need to address the colonial history of science and contact in order to ethically explore space and seek to contact extraterrestrial life. \citet{Lempert2021} explained how efforts which begin with the purest of scientific intentions can serve imperialist interests in a culture of settler colonialism, drawing connections between the Endeavour voyage to study the transit of Venus in 1768, which ultimately led to severe colonial violence, and SETI.  \citet{Charbonneau2021} addressed SETI's simultaneous embrace and unease with historical projections of contact and its implications, exploring SETI as a reflection of humanity and how it perceives itself and emphasizing the importance of incorporating social sciences into SETI work. 

In a less SETI-focused portion of the AICRJ issue, \citet{Maile2021} described the struggle for sovereignty between indigenous people and colonizers in Hawaii, particularly with respect to the ongoing Thirty Meter Telescope project and protests on Mauna Kea. \citet{Painter2021} explored the impact of jurisdiction on potential contact with extraterrestrial life on Earth, with a historical look at the FBI's focus on indigenous land during the cow mutilations of the 1970s. \citet{Kite2021} contrasted the fear of the unknown and conspiracy theory culture ingrained in the colonial United States with respect to the extraterrestrial and supernatural with the embracing of the unknown in indigenous Lakota culture. \citet{Tallbear2021} creatively depicted the relationships between indigenous studies and settler-based fields and sciences.

\subsection{Anthropocentrism}

Several papers this year discussed the anthropocentric assumptions involved in SETI, proposing alternative ways of considering extraterrestrial life and communication. \citet{Melka2021} described the difficulty humans have in interpreting our own messages and histories, as well as in communication with other Earth-based life. They argued that planning for contact with ETI requires semiotic/linguistic and psychological perspectives. In a similar vein, \citet{DOBLER2021} proposed a redefining of ``exopsychology,'' a subfield of psychology focused on extraterrestrial life and their human representation, acknowledging the anthropocentric biases in SETI and proposing methods to mitigate them. \citet{Traphagan2021b} critiqued the notion of linear progress and societal evolution that exists in SETI.

\subsection{Contact \& Messaging Extraterrestrial Intelligence}

As always, METI occupied a large part of the discussion in 2021. \citet{Kerins2021} explored mutual detectability as a solution to the ``SETI paradox,'' which notes the contradiction between Earth's search for intentional SETI signals with our hesitance to participate in METI, from a game theory perspective. They concluded that the onus to initiate communication lies with the more detectable civilization, or the habitable zone planet around the smallest host star, with is typically not Earth, motivating a SETI search targeting Earth-analog transiting planets around subsolar luminosity hosts in the Earth Transit Zone. 

\citet{Hatfield2021} explored how and by whom SETI and METI decisions should be made. They conclude that scientists, including the social sciences and the humanities, should lead on technical scientific issues and how to communicate, while questions of judgement, about whether action should be taken and what to say, must involve representatives of broader society from all around the world.

Rather than focusing on whether METI should be performed, \citet{Santana2021} questioned whether an initial METI message should be peaceful or belligerent. Arguing that extraterrestrial life need not be similar to Earth's, they concluded that defensive belligerence is the most logical. % lol what
\citet{Traphagan2021} also explored the content of METI messages, in particular those which have already been sent out. They noted that messages such as the Golden Records could be interpreted incorrectly or as being intentionally deceptive, putting humanity at risk, and concluded by suggesting the development of review boards  to control the content of government-sponsored METI. % got interlibrary loan - can take deeper look to make sure i got the key message

\citet{Bohlander2021} explored the topic of ``metalaw,'' or legal rules regarding the relationships between intelligent life on Earth and that elsewhere in the universe, in the context of our limited human viewpoint. 

SETI/METI were also the primary topics of two books and several other book chapters this year. \citet{McConnell2021} explored the science of interstellar communication, detailing how the contact process may unfold, what information could be sent, and how humanity will participate in the interpretation effort. This follows \citet{McConnell2020}, who proposed an architecture inspired by the Deep Space Network for data sharing in the event of a SETI detection (which should have been included in the SETI in 2020 paper list, an oversight which we correct here).

\citet{green21} explored ethics as they apply to space exploration and use, including an in-depth discussion of METI. %based on the book summary and what manasvi said - i have requested interlibrary loan
Published in 2021, \emph{Astrobiology: Science, Ethics, and Public Policy} included several discussions of METI and contact. \citet{Peters2021} analyzed the ethical considerations for engagement with ETIs lesser, equal, and superior in intelligence to humans, emphasizing the need to respond with care to any form of life. \citet{Haramia2021} provided an ethical assessment of whether and how SETI and METI should be performed, arguing for a planetocentric and ecosystemic approach to making these decisions. \citet{Rappaport2021} explored how humanity may assess the presence of ethics in ETI in different contact scenarios.

\citet{Cakir2021} discussed legal considerations regarding contact with ETI, in the context of the Mars Agreement Including Human Settlements. Though more focused on accidental messages to ETI than intentional, \citet{Baxter2021} explored the detectability of ``Big History'', or humanity's past, present, and future. \citet{Garrett2021} contrasted the limited view of Big History on humanity's future with the wide perception of life and intelligence offered by SETI.

\citet{dePaulis21} presented {\it COGITO in Space}, an art project which is sending human brain activity signals in the form of radio waves into space, with a goal more focused on reflecting on the relationships between humans and space than performing METI.

\subsection{History}
Lastly, two papers focused on aspects of SETI history. \citet{Smith2021a} discussed the history of the ``Fermi Paradox'' and a similar line of questioning in Jules Verne's 1869 novel \emph{Around the Moon}. \citet{Gray2021} described Project Ozma II, which observed 670 nearby stars for radio signals between 1972 and 1976, and its context as a major milestone for SETI.

\section{Looking Ahead}

While 2022 began with the cancellation of the 239\th\ meeting of the American Astronomical Society meeting, which was set to feature an oral presentation session on technosignatures, we can look forward to many in-person and hybrid SETI gatherings in the coming year. We are particularly excited for the first Penn State SETI Symposium set for June 2022, after being postponed twice. Additionally, we look forward to SETI sessions at several larger conferences throughout the year: the Astrobiology Science Conference in Atlanta/online in May, the Committee on Space Research Assembly in Athens/online in July, and the International Astronautical Congress in Paris in September.

With only two years of data, it is impossible to identify trends in the SETI literature, but we can optimistically look forward to an increase in the field's research output. We look forward to a third installment of this series with an update on all of these items in a year or so, and of course we can all follow the field as it develops in the literature at the \href{https://seti.news}{seti.news} website.

\begin{acknowledgements}

We thank Manasvi Lingam and Chelsea Haramia for helpful comments and for providing several references we had missed. 
We thank Virginia Trimble for her APxx series, and Clement Vidal for his encouragement and ideas for this paper. % and for her suggestion that I owe her a bottle of vermouth because of this paper, which I choose to interpret as an implicit endorsement (n.b.\ the debt has been paid). I thank the anonymous referees for their reviews. 
We thank J\"urgen Lehman for sharing a list of relevant papers missing from the SETI library.

This research has made use of NASA's Astrophysics Data System Bibliographic Services. This research was supported by the Center for Exoplanets and Habitable Worlds and the Penn State Extraterrestrial Intelligence Center, which are supported by the Pennsylvania State University and the Eberly College of Science.

\end{acknowledgements}

\bibliography{refs}

\end{document}